\documentclass[aps,prd,12point,twocolumn,nofootinbib,showpacs,superscriptaddress]{revtex4-2}
\def\theequation{\arabic{section}.\arabic{equation}}
\usepackage{psfrag}
\usepackage{subfigure}
\usepackage{color}
\usepackage{mathrsfs}
\usepackage{graphicx}
\usepackage{amssymb, bm}
\usepackage{amsmath, amsthm}
\usepackage{epstopdf}
\usepackage{hyperref}
\usepackage{enumerate}
\usepackage{longtable}

\usepackage{amsfonts}
\usepackage[T1]{fontenc}
\usepackage{bm}
\usepackage{amsmath}  
\usepackage{graphicx} 
\usepackage{color}
\usepackage{appendix}

\newcommand{\be}{\begin{equation}}
\newcommand{\ee}{\end{equation}}

\begin{document}
\def\theequation{\arabic{section}.\arabic{equation}} 
%\preprint{}

\title{Brans-Dicke analogue of the Roberts geometry}

\author{Bardia H. Fahim}
\email[]{bardia.fahim@usask.ca}
%\thanks{}
\affiliation{Department of Physics \& Astronomy, Bishop's University, 
2600 College Street, Sherbrooke, Qu\'ebec, 
Canada J1M~1Z7}

\author{Valerio Faraoni}
\email[]{vfaraoni@ubishops.ca}
%\thanks{}
\affiliation{Department of Physics \& Astronomy, Bishop's University, 
2600 College Street, Sherbrooke, Qu\'ebec, 
Canada J1M~1Z7}

\author{Andrea Giusti}
\email[]{agiusti@ubishops.ca}
%\thanks{}
\affiliation{Department of Physics \& Astronomy, Bishop's University, 
2600 College Street, Sherbrooke, Qu\'ebec, 
Canada J1M~1Z7}


%\date{\today}

\begin{abstract}

We report a new one-parameter family of spherically symmetric, 
inhomogeneous, and time-dependent solutions of the vacuum Brans-Dicke 
field equations which are conformal to the Roberts scalar field geometries 
of Einstein gravity. The new solution is spherical and time-dependent and 
contains a naked central singularity. We use it as a seed to 
generate another two-parameter family of solutions using  a known symmetry 
of  vacuum Brans-Dicke gravity.

\end{abstract}

\pacs{}
% insert suggested keywords - APS authors don't need to do thisW
%\keywords{}

\maketitle

\section{Introduction}
\setcounter{equation}{0}
\label{sec:1}

Einstein's theory of General Relativity (GR) conflicts with quantum 
mechanics, as demonstrated by the fact that attempts to quantize GR 
produce, in the low-energy limit, theories of gravity which deviate 
from 
GR due to extra degrees of freedom or higher order terms in their field 
equations. From the theoretical point of view, therefore, GR must be 
modified in order to make it compatible with quantum field theory. A more 
urgent motivation to explore gravity beyond Einstein theory comes from 
observational cosmology. If the present acceleration of the universe 
discovered in 1998 with type Ia supernovae is to be explained within the 
realm of GR, one needs an incredibly fine-tuned cosmological constant 
$\Lambda$ or a completely {\em ad hoc} dark energy sourcing the Einstein 
equations. Instead of postulating these rather embarassing and  
{\em ad hoc} ingredients of the standard $\Lambda$-Cold Dark Matter 
($\Lambda$CDM) cosmological model, a viable alternative consists of 
modifying gravity at large scales, while preserving GR at small scales. 
Moreover, while the linear approximation to GR is well tested at Solar 
System scales and has received a spectacular confirmation with the {\em 
LIGO} detections of gravitational waves \cite{LIGO1,LIGO2,LIGO3,LIGO4}, 
the theory is not tested at most spatial scales and in most curvature 
regimes \cite{Bertietal2015,Psaltis}. Currently, the most popular 
alternative to GR to explain the cosmic acceleration without invoking dark 
energy is probably the class of $f({\cal R})$ theories of gravity, where 
${\cal R}$ is the Ricci scalar of the metric connection \cite{CCT} (see 
\cite{reviews} for reviews). $f({\cal R})$ theories are nothing but 
scalar-tensor theories in disguise with vanishing Brans-Dicke coupling 
$\omega$ and equipped with a complicated potential for the scalar degree 
of freedom $f'({\cal R})$ \cite{reviews}.

Brans-Dicke theory \cite{BD} is the prototypical alternative to GR 
and the simplest representative of scalar-tensor gravity \cite{ST}. 
Analytical solutions of the field equations are useful to understand 
the physics of this theory and of its scalar-tensor generalizations. 
Inhomogeneous and 
time-dependent scalar field solutions are rare both in Einstein and in 
alternative gravity and it is interesting to expand the meagre catalogue 
available. One example of spherically symmetric, inhomogeneous, and 
time-dependent solution of the Einstein equations is the Roberts geometry 
sourced by a massless, minimally coupled scalar field\footnote{A version 
with a conformally coupled scalar has also been proposed 
\cite{Robertsconformal}.} \cite{Roberts}. The Roberts solution has been 
used as an example in the study of critical phenomena occurring during the 
gravitational collapse of scalar fields \cite{Choptuik93, Brady94, 
Oshiro94, HMN}, is of interest for Cosmic Censorship \cite{Roberts} and 
wormhole formation 
\cite{Oliveira96,WangOliveira97,Frolov00,Maeda09,Maeda15,Almeida16}, and 
has been the 
subject of some attention due to the difficulty of obtaining dynamical 
and inhomogeneous analytical solutions of the Einstein equations 
\cite{Burko97,Maeda09,Maeda15}.

The Roberts geometry \cite{Roberts} (with an error corrected in 
Refs.~\cite{Sussman91,Brady94,Oshiro94,Burko97, 
Hayward00,ClementHayward01}) is a spherical, 
continuously 
self-similar spacetime since it admits a homothetic Killing vector ﬁeld 
$\xi^c$ satisfying $\pounds_{\xi} g_{ab}= 2 \lambda g_{ab}$, where 
$\lambda$ is a constant \cite{Roberts}. Its 
stability was studied in \cite{Frolov97,Frolov99}. Here we 
adopt\footnote{We follow the notations of Ref.~\cite{Waldbook}.}  the  
corrected and slightly generalized Roberts solution given by Burko in 
double null coordinates $\left(u,v, \vartheta, \varphi \right)$ 
\cite{Burko97},
\be
d\tilde{s}^2 =-dudv +r^2 (u,v) d\Omega_{(2)}^2 \,, \label{Robertsmetric}
\ee
\be
r^2 \left( u,v \right) = 
\frac{1}{4} \left[ \left(1-4\sigma^2 \right) v^2 -2uv +u^2 \right] 
\,,\label{rsquared}
\ee
where $d\Omega_{(2)}^2 \equiv d\vartheta^2 +\sin^2 \vartheta \, 
d\varphi^2$ is the line element on the unit 2-sphere and the scalar field 
is
\be
\tilde{\phi} =\pm \frac{1}{2} \ln \left[ \frac{(1-2\sigma)v 
-u}{(1+2\sigma)v -u } \right] \,,\label{Robertsscalar}
\ee
where $\sigma $ is a constant. In order for the areal radius 
squared~(\ref{rsquared}) and for the argument of the logarithm in 
Eq.~(\ref{Robertsscalar}) to be positive, it must be $|\sigma| <  
1/2$. In the limit $\sigma \rightarrow 0$ the scalar 
$\tilde{\phi}$ disappears and one recovers Minkowski spacetime.  We also 
note that the right hand side of 
Eq.~(\ref{rsquared}) can be written as
\be
r^2 \left( u,v \right) 
=  \frac{1}{4} \left[ (1-2\sigma)v -u \right] \left[ (1+2\sigma) v -u 
\right]  \,,\label{previous}
\ee
which will be useful later. Therefore, the origin $r=0$ corresponds to 
$u=\left( 1 \pm 2\sigma\right)v$. Since $r^2\geq 0$, Eq.~(\ref{previous}) 
guarantees that the argument of the logarithm in the Roberts scalar 
field~(\ref{Robertsscalar}) is positive (the absolute value of 
this argument appearing in this equation in~\cite{Burko97} is redundant).  
In the region 
in which the gradient of the scalar field is timelike, the Roberts 
solution reproduces \cite{Burko97} a 1967 stiff fluid solution of Gutman 
and Bespalko \cite{GutmanBespalko}.

The Einstein equations reduce to 
\be
\tilde{R}_{ab}= 8\pi \nabla_a \tilde{\phi} \nabla_b \tilde{\phi} 
\,,
\ee
where $ \tilde{R}_{ab} $ is the Ricci tensor, while the Ricci scalar is
\be
\tilde{R}=8\pi  \tilde{g}^{ab}\nabla_a \tilde{\phi} \nabla_b 
\tilde{\phi} 
= - 32 \pi \tilde{\phi}_{,u} \tilde{\phi}_{,v} = \frac{ 
8\pi \sigma^2 uv}{r^4} 
\ee
and diverges as $r\rightarrow 0$, where there is a spacetime singularity. 
This singularity is not covered by apparent horizons, whose radii would be 
the 
positive roots of the equation $\tilde{g}^{ab} \nabla_a r \nabla_b r=0$. 
For the Roberts solution, this equation reads
\be
\tilde{g}^{ab} \nabla_a r \nabla_b r =-\frac{1}{4 \, r^2} \left[ 
(1-4\sigma^2)v-u \right] (u-v) =0
\ee
and the only roots are $v=u$ or $v=\frac{u}{1-4\sigma^2} $ (remember that 
$|\sigma|<1/2$). But, using 
Eq.~(\ref{rsquared}),  $v=u$ would imply $r^2 =-\sigma^2 u^2<0$, which 
is impossible. Similarly, $v=u/(1-4\sigma^2)$ would imply 
$r^2=-\frac{\sigma^2 u^2}{1-4\sigma^2}<0$, which is also impossible (the 
exceptions are $u=v=0$ which corresponds to $r=0$ and to the absence of  
apparent horizon positive roots). Therefore, there are no apparent 
horizons and the central singularity is naked. 
The Misner-Sharp-Hernandez mass $ M_\text{MSH}$ is defined 
by $
1-2M_\text{MSH}/r=\tilde{g}^{ab}\nabla_a r \nabla_b r $ \cite{MSH,  MSH-us}, 
which gives
\be
M_\text{MSH} = \frac{r}{2} \left( 1- \tilde{g}^{ab}\nabla_a r \nabla_b r 
\right) = -\frac{\sigma^2 \, uv}{2r} \,.
\ee
It is negative in the entire region $uv>0$ forming the past and future 
light cone of the origin $r=0$.

In the following, we regard the Roberts scalar field solution of GR as the 
Einstein frame version of a Brans-Dicke counterpart, which constitutes a 
new solution explored in this paper. Analytical solutions of 
scalar-tensor gravity with the gravitational scalar field $\phi$ 
propagating along null fronts are unknown, except for certain exact plane 
waves---some of them are obtained by matching spacetime regions along 
null shells \cite{Bressange,BarrabesBressange} and others are exotic 
stealth wave solutions of Brans-Dicke-Maxwell gravity \cite{Robinson}. It 
is interesting, therefore, to examine the Brans-Dicke analogue of the 
Roberts solution and, possibly, similar solutions of $f({\cal R})$ 
gravity.

Let us recall some basics: the Jordan frame 
action of vacuum  Brans-Dicke theory is \cite{BD}
\be \label{BDaction}
S_\text{BD} = \int d^4x \, \frac{ \sqrt{-g}}{16 \pi}
\left(\phi  \mathcal{R}-\frac{\omega}{\phi} \nabla^c\phi
\nabla_c\phi -V(\phi) \right) \,,
\ee
where $\phi$ is the gravitational scalar field (approximately 
equivalent to the inverse of the gravitational coupling 
$G_{\text{eff}}$), $V(\phi)$ is the scalar field potential, and $\omega$ 
is the 
constant ``Brans-Dicke coupling'' parameter, while $g$ is the 
determinant of the spacetime  metric $g_{ab}$. By varying this  
action, one obtains the vacuum Brans-Dicke field equations \cite{BD}
\begin{eqnarray}
R_{ab}-\frac{{\cal R}}{2} g_{ab} &=& \frac{\omega}{\phi^2}
\left( \nabla_a\phi \nabla_b\phi -\frac{1}{2}\, g_{ab}
\nabla^c \phi\nabla_c\phi \right)  \nonumber\\
&&\nonumber\\
&\, & +\frac{1}{\phi} \left( \nabla_a\nabla_b\phi -g_{ab} 
\Box \phi 
\right) -\frac{V}{2\phi}\, g_{ab} \,,\label{BDfe} 
\nonumber\\
&&\\
\Box \phi  &=& \frac{1}{2\omega+3} \left( \phi \, 
\frac{dV}{d\phi} 
-2V \right) \,. \label{box}
\end{eqnarray}

A second representation of scalar-tensor gravity, the 
Einstein conformal frame \cite{Dicke} is obtained by means of the 
conformal  transformation of the  metric
\be \label{metric transformation}
g_{ab} \rightarrow \tilde{g}_{ab} = \phi \, g_{ab}
\ee
and the scalar field redefinition
\be
\phi \rightarrow \tilde{\phi}=\sqrt{\frac{|2\omega+3|}{16
\pi}} \, \ln \left( \frac{\phi}{\phi_{*}}\right) \,,
\ee
where $\phi_{*}$ is a constant and $\omega\neq -3/2$. In terms of these 
new variables, the  vacuum 
Brans-Dicke action~(\ref{BDaction}) assumes the Einstein frame form
\be
S_\text{BD} = \int d^4x\sqrt{-\tilde{g}}\left[
\frac{\tilde{\mathcal{R}}}{16 \pi}-\frac{1}{2} \, 
\tilde{g}^{ab}\nabla_a\tilde{\phi} \nabla_b\tilde{\phi}  
-\tilde{V}(\tilde{\phi}) \right]  \,, \label{BDactionEframe}
\ee
where
\be
\tilde{V}(\tilde{\phi}) = \frac{V(\phi)}{\phi^2}\left|_{\phi=\phi( 
\tilde{\phi}) } 
\right. \,. \label{ppotential} 
\ee
We label Einstein frame quantities with a tilde. 
Formally,~(\ref{BDactionEframe}) is  
the Einstein-Hilbert action coupled to a standard matter scalar 
field. The vacuum field equations in the Einstein frame  are
\begin{eqnarray}
\tilde{R}_{ab}-\frac{1}{2} \, \tilde{g}_{ab} \tilde{
{\cal R}} &=& 8\pi \left( \nabla_a \tilde{\phi} \nabla_b
\tilde{\phi} -\frac{1}{2} \, \tilde{g}_{ab}
\tilde{g}^{cd} \nabla_c \tilde{\phi}\nabla_d \tilde{\phi}
\right)  \nonumber\\
&&\nonumber\\
&\, & -\tilde{V}(\tilde{\phi}) \tilde{g}_{ab} \,, \label{Eframefe}
\end{eqnarray}
\be
 \tilde{g}^{ab} \tilde{ \nabla}_a  \tilde{ \nabla}_b
\tilde{\phi} -\frac{d\tilde{V}}{d\tilde{\phi}} = 0 \,. 
\label{EframeKG}
\ee
Given a solution of the Einstein equations sourced by a minimally coupled 
scalar field, we can interpret it as the Einstein frame counterpart of a 
Jordan frame Brans-Dicke gravity and map it to its representation in the 
Jordan frame. As a {\em caveat}, in general the Jordan frame scalar field 
potential $V(\phi)$ obtained from a reasonable Einstein frame 
potential $\tilde{V}( \tilde{\phi})$ is physically unmotivated, 
but this fact will not be of concern here since the Roberts solution that 
we consider has zero potential and this property transfers to the Jordan 
frame, as is well known \cite{NMCrefs, CMB}.

Now, on to $f({\cal R})$ gravity: metric $f({\cal R})$ theories 
of gravity 
are a subclass of Brans-Dicke gravity with action \cite{reviews} 
\be
S = \int d^4 x \, \frac{\sqrt{-g}}{16\pi}  \, f({\cal R})   
\label{f(R)action}
\ee
{\em in vacuo}, where $f({\cal R})$ is a non-linear function of the Ricci 
scalar ${\cal R}$. By introducing the new scalar field  $\phi \equiv 
f'({\cal 
R})$ with potential 
\be \label{f(R)potential}
V(\phi)= \phi {\cal R}(\phi) -f\left( {\cal R}(\phi)  
\right) \,,
\ee
one shows \cite{reviews} that the action~(\ref{f(R)action}) is equivalent 
to the vacuum Brans-Dicke action 
\be
S = \int d^4 x \, \frac{ \sqrt{-g}}{16\pi}  \left[ \phi 
{\cal 
R}-V(\phi) \right]  \,,
\ee
with Brans-Dicke parameter $\omega=0$ and the 
potential~(\ref{f(R)potential}).

\section{A new solution of vacuum Brans-Dicke theory}
\setcounter{equation}{0}
\label{sec:2}

We now regard the Roberts spacetime $\left( 
\tilde{g}_{ab}, \tilde{\phi} 
\right)$ as the Einstein frame version of a Brans-Dicke solution $\left( 
g_{ab}, \phi \right)$, which we map back to the Jordan conformal frame. 
Their relation is
\begin{equation}
    ds^2=\phi^{-1}d\Tilde{s}^2 \,,
\end{equation}
\begin{equation}
    \phi=\phi_* \exp\left[  \sqrt{\frac{16\pi}{\left| 2\omega+3 \right|} } 
\, \Tilde{\phi} \right] \,.
\end{equation}
Applying these on the Roberts 
solution~(\ref{Robertsmetric})-(\ref{Robertsscalar})  yields
\begin{equation}
ds^2=  \left[ \frac{ (1-2\sigma)v-u }{ (1+2\sigma)v-u } \right]^{
\mp 2 \sqrt{ \frac{\pi}{|2\omega+3|} }  
} 
\left[ -dudv + r^2(u,v) d\Omega^2_{(2)}  \right] \,, \label{new1}
\end{equation}
\begin{equation}
\phi=\phi_* \left[ \frac{(1-2\sigma)v-u}{ (1+2\sigma)v-u} \right]^{\pm 2 
\sqrt{\frac{\pi}{ \left|  2\omega+3\right| }}} \,,\label{Jframescalar}
\end{equation}
where $\phi_*>0$ (this constant is omitted without consequences in the 
expression of $ds^2$) and $r$ is given by Eq.~(\ref{rsquared}).

The Jordan frame areal radius is
\begin{eqnarray}
R(u,v) &=& \frac{r(u,v)}{ \sqrt{\phi} } =
\frac{1}{2} \left[ \frac{ (1-2\sigma)v-u }{(1+2\sigma)v-u }\right]^{\mp 
\sqrt{ \frac{\pi}{|2\omega+3|}} } \nonumber\\
&&\nonumber\\
&\, & \times \sqrt{  (1-4\sigma^2)v^2-2uv+u^2} 
\end{eqnarray}
where we used Eq.~(\ref{rsquared}), from which it follows that the Jordan 
frame origin $R=0$ corresponds to $r=0$ or to $ u= \left(1 \pm 2\sigma 
\right)v $, which gives again $r=0$, therefore there is a one-to-one 
correspondence between $R=0$ and $r=0$. The Jordan frame 
scalar~(\ref{Jframescalar}) diverges at the origin $R=0$.

When they exist, the apparent horizons of a spherically symmetric metric 
are located by the roots of the equation $\nabla^c R \nabla_c R=0$, which 
here takes the form
\begin{eqnarray}
\nabla^c R \nabla_c R &=& 2g^{u v}R_{,u}R_{,v}=-4\phi R_{,u}R_{,v} 
\nonumber\\
&&\nonumber\\
&=& - 4 \left(r_{,u} -\frac{r\phi_{,u}}{2\phi} \right)   
\left(r_{,v}-\frac{r\phi_{,v}}{2\phi} \right)   \,,\label{misonrotto}
\end{eqnarray}   
where we used the inverse metric
\be
\left( g^{\mu\nu} \right) = \left( \begin{array}{cccc}
0 & -2\phi & 0 & 0 \\
-2\phi & 0 & 0 & 0 \\
0 & 0 & \frac{\phi}{r^2}  & 0  \\
0 & 0 & 0 & \frac{\phi}{r^2\sin^2\vartheta} \\
\end{array} \right) \,. \label{inversemetric}
\ee
We must recover  Minkowski space in the limit $\sigma \rightarrow 0$ in 
which $\phi \rightarrow $~const. and $
ds^2 \rightarrow -dudv+r^2(u,v) d\Omega_{(2)}^2 $. Then, $R$ depends on 
both $u$ and $v$ in an essential way and this is true also for $\sigma\neq 
0$, hence the vanishing of $R_{,u}$ or $R_{,v}$ does not make sense 
physically. We conclude that there are no roots of Eq.~(\ref{misonrotto}) 
and no apparent horizons.

The Jordan frame Ricci scalar is obtained from the trace of 
Eq.~(\ref{BDfe}) 
\begin{equation}    
{\cal R} =\frac{\omega}{\phi^2}\nabla^a\phi\nabla_a\phi 
+\frac{3\Box\phi}{\phi}+\frac{2\Tilde{V}}{\phi},
\end{equation}
where the second and third terms in the right hand side are zero, as 
$\phi$ is a free scalar field that satisfies $\Box\phi=0$ (cf. 
Eq.~(\ref{box})). The Ricci scalar becomes 
\begin{eqnarray}
{\cal R} &=& \frac{\omega}{\phi^2} \, g^{ab} \nabla_a \phi \nabla_b \phi 
=\frac{2\omega}{\phi^2} \, g^{uv} \partial_u \phi \partial_v \phi 
\nonumber\\
&&\nonumber\\
&=&  \frac{16\pi \omega \sigma^2}{\mid 2\omega+3 \mid}\frac{\phi \, 
uv}{r^4} =
\frac{16\pi \omega \sigma^2}{\mid 2\omega+3 \mid}\frac{uv}{\phi R^4}  
\,,\label{JframeR}
\end{eqnarray}
where we used the inverse metric~(\ref{inversemetric}) and 
\begin{eqnarray}
\partial_u \phi &=& \mp 2\sigma \sqrt{ \frac{\pi}{|2\omega+3|} } \, 
\frac{\phi v}{r^2} 
= \mp 2\sigma \sqrt{ \frac{\pi}{|2\omega+3|} } \, 
\frac{v}{ R^2} \,,\nonumber\\
&&\\
\partial_v \phi &=& 
\pm 2\sigma \sqrt{ \frac{\pi}{|2\omega+3|}} \, \frac{\phi u}{ r^2} 
= \pm 2\sigma \sqrt{ \frac{\pi}{|2\omega+3|}} \, \frac{ u}{ R^2} 
\,.\nonumber\\
&&
\end{eqnarray}
When $\omega\neq 0$, the Ricci scalar diverges as $ R \rightarrow 0$ 
together with $\phi$ and there exists a naked central singularity. When 
$\omega=0$, the Ricci scalar vanishes identically, but the Brans-Dicke 
scalar field $\phi$ still diverges at the origin with the 
Kretschmann 
scalar ${\mathcal K}\equiv R_{abcd}R^{abcd}$. This is given, for general 
values of $\omega$, by

%\begin{widetext} 
%{\color{red} With the $+$:
%\begin{equation}
%\begin{split}
%\mathcal{K}_{+}=
%\frac{4 \, \sigma^2 \, \phi^2}{\left| 2 \omega +3\right| ^2 \, r^8} 
%\Big\{&4 \sqrt{\pi} \sigma \, u \, v \, \left| 2 \omega +3\right| ^{3/2} 
%\left[\left(4 \sigma^2-1\right) v^2+u^2\right] 
%+3 \, \sigma^2 \, u^2 \, v^2 \, \left( 2 \omega +3\right)^2\\
%&+2 \pi \left| 2 \omega +3\right|  \left[2 \left(10 \sigma^2-3\right) u^2 
%\, v^2+2 \left(1-4 \sigma^2\right) u \, v^3+\left(1-4 \sigma^2\right)^2 
%v^4 +2 u^3 \, v + u^4\right]\\
%&+16 \pi ^{3/2} \sigma u \, v \sqrt{\left| 2 \omega +3\right|} 
%\left[\left(4 \sigma^2-1\right) v^2+u^2\right]+48 \pi ^2 \sigma^2 \, u^2 
%\, v^2\Big\}
%\end{split}
%\end{equation}
%With the $-$:
%\begin{equation}
%\begin{split}
%\mathcal{K}_{-}=
%\frac{4 \, \sigma^2 \, \phi^2}{\left| 2 \omega +3\right| ^2 \, r^8} 
%\Big\{&-4 \sqrt{\pi} \sigma \, u \, v \, \left| 2 \omega +3\right| ^{3/2} 
%\left[\left(4 \sigma^2-1\right) v^2+u^2\right]
%+3 \sigma^2 \, u^2 \, v^2 \left( 2 \omega +3\right)^2\\
%&+2 \pi  \left| 2 \omega +3\right|  \left[2 \left(10 \sigma^2-3\right) 
%u^2 \, v^2+2 \left(1-4 \sigma^2\right) u \, v^3
%+\left(1-4 \sigma^2\right)^2 v^4+2 \, u^3 \, v+u^4\right]\\
%&-16 \pi ^{3/2} \, \sigma \, u \, v \sqrt{\left| 2 \omega +3\right| } 
%\left[\left(4 \sigma ^2-1\right) v^2+u^2\right]
%+48 \pi ^2 \sigma^2 \, u^2 \, v^2\Big\}
%\end{split}
%\end{equation}
%}

\begin{widetext}
\begin{equation}
\begin{split}
\mathcal{K} = 
\frac{4 \, \sigma^2 \, \phi^2}{\left| 2 \omega +3\right| ^2 \, r^8} 
\Big\{&\pm 4 \sqrt{\pi} \sigma \, u \, v \, \left| 2 \omega +3\right| ^{3/2} \left[\left(4 \sigma^2-1\right) v^2+u^2\right] 
+3 \, \sigma^2 \, u^2 \, v^2 \, \left( 2 \omega +3\right)^2\\
&+2 \pi \left| 2 \omega +3\right|  \left[2 \left(10 \sigma^2-3\right) u^2 \, v^2+2 \left(1-4 \sigma^2\right) u \, v^3+\left(1-4 \sigma^2\right)^2 v^4 +2 u^3 \, v + u^4\right]\\
&\pm16 \pi ^{3/2} \sigma u \, v \sqrt{\left| 2 \omega +3\right|} \left[\left(4 \sigma^2-1\right) v^2+u^2\right]+48 \pi ^2 \sigma^2 \, u^2 \, v^2\Big\}
\end{split}
\end{equation}
\end{widetext}

\section{``Conformal Roberts'' is not a solution of $f({\cal R})$ gravity}
\setcounter{equation}{0}
\label{sec:3}

Let us consider now the possibility that the new Brans-Dicke solution is 
also a solution of $f({\cal R})$ gravity. Since there is no potential, it 
must be (cf. Eq.~(\ref{f(R)potential})) 
\be
V(\phi)= \phi {\cal R}-f({\cal R})=0
\ee
which integrates to $f({\cal R})= f_0 \, {\cal R}$, where 
$f_0$ is a constant. In addition, it 
must 
be $\omega=0$ and then the putative solution becomes
\be
ds_{(0)}^2 =\left[ \frac{(1-2\sigma)v-u}{(1+2\sigma)v-u} \right]^{ \mp2 
\sqrt{\pi/3}} \left[ -dudv+r^2(u,v) d\Omega_{(2)}^2 \right] 
\,,\label{accidenti1}
\ee
\be
\psi=\phi_*  \left[ \frac{(1-2\sigma)v-u}{(1+2\sigma)v-u}  \right]^{ 
\pm 2 \sqrt{ \pi/3}} \,,\label{accidenti2}
\ee
and it would seem that these expressions could provide a solution of GR. 
This is not true because, according to Eq.~(\ref{JframeR}), $\omega=0 $ 
also implies ${\cal R}=0$ and, in GR, it is instead 
${\cal R} \propto \nabla^c \psi \nabla_c 
\psi$ which, in general, is incompatible with ${\cal R}=0$. 

We conclude that the vacuum Brans-Dicke geometry (\ref{accidenti1}), 
(\ref{rsquared}), and (\ref{accidenti2}) is not a solution of $f({\cal 
R})$ gravity nor of the Einstein-Klein-Gordon equations.

\section{Another two-parameter family of solutions}
\setcounter{equation}{0}
\label{sec:4}

Vacuum Brans-Dicke theory is invariant under the operation 
\cite{myBDlimit}
\begin{eqnarray}
g_{ab} & \rightarrow \hat{g}_{ab} = \phi^{2\alpha} 
g_{ab} \,, \label{symmetry1}\\
&&\nonumber\\
\phi & \rightarrow \hat{\phi} = \phi^{1-2\alpha} \,, \label{symmetry2}
\end{eqnarray}
for $\alpha \neq 1/2$ (the conformal transformation of the 
metric~(\ref{symmetry1}) has nothing to do with the conformal map relating 
Jordan and Einstein frames). A hat denotes geometric quantities 
constructed with the conformally rescaled metric $\hat{g}_{ab}$. The  
well-known transformation properties under the map $ 
\hat{g}_{ab}=\Omega^2 g_{ab}$ \cite{Synge, Waldbook, Carroll, mybook} 
\begin{eqnarray}
\hat{g}^{ab} &=& \Omega^{-2} g^{ab} \,,\\
&&\nonumber\\
\sqrt{-\hat{g}} & = & \Omega^4  \sqrt{-g} \,,\\
&&\nonumber\\
\hat{R} &=& \Omega^{-2} \left( R-\frac{6\Box\Omega}{\Omega}\right) \,,
\end{eqnarray}
plus the use of Eq.~(\ref{symmetry2}) yield
\begin{eqnarray}
R &=& \phi^{2\alpha} \hat{R}  
- \frac{6\alpha ( 1-\alpha)}{(1-2\alpha)^2} \,  
 \phi^{6\alpha-2} \hat{g}^{ab} 
\hat{\nabla}_a \hat{\phi} \hat{\nabla}_b \hat{\phi} \nonumber\\
&&\nonumber\\
&\, & +\frac{6\alpha}{1-2\alpha} \, 
\phi^{4\alpha-1}\hat{\Box}\hat{\phi} \,. \label{mmminch}
\end{eqnarray}
The d'Alembertian in the right hand side of Eq.~(\ref{mmminch}) can be 
written as 
\be
\frac{6\alpha}{1-2\alpha} \,\sqrt{-\hat{g}} \,  
\hat{\Box}\hat{\phi}= 
\frac{6\alpha}{1-2\alpha} \, \partial_{\mu} \left( \sqrt{-\hat{g}} 
\,\hat{g}^{\mu\nu} \partial_{\nu} \hat{\phi} \right) \,,
\ee
and is integrated to a boundary term when placed in the action 
integral. By dropping this term (which is irrelevant in the variation 
leading to the Brans-Dicke field equations), the vacuum Brans-Dicke 
action~(\ref{BDaction}) without potential reads 
\begin{eqnarray}
S_\text{BD} &=& \int d^4x \sqrt{ -\hat{g}} \left\{ \hat{\phi} \hat{R} 
-\left[ \frac{\omega}{(1-2\alpha)^2}+ 
\frac{6\alpha(1-\alpha)}{(1-2\alpha)^2} \right] \right. \nonumber\\
&&\nonumber\\
&\, & \left. \frac{ \hat{g}^{ab}}{\hat{\phi}} \, \hat{\nabla}_a \hat{\phi} 
\hat{\nabla}_b \hat{\phi}  \right\} \,.
\end{eqnarray}
If we rename the Brans-Dicke coupling as
\be
\hat{\omega}( \omega, \alpha) =  \frac{ \omega 
+6\alpha(1-\alpha)}{(1-2\alpha)^2} \,, \label{newomega}
\ee
then the action reads \cite{myBDlimit} 
\be
S_\text{BD}= \int d ^4x \sqrt{ -\hat{g}} \left[ \hat{\phi} \hat{R} 
- \frac{ \hat{\omega}}{\hat{\phi}} \,  
\hat{g}^{ab}  \hat{\nabla}_a \hat{\phi} 
\hat{\nabla}_b \hat{\phi} \right] \,,
\ee
which is again of the Brans-Dicke form. Therefore, the 
operation~(\ref{symmetry1}), (\ref{symmetry2}), and~(\ref{newomega}) is a 
symmetry of vacuum Brans-Dicke theory. It can be shown that, as $\alpha $ 
varies, it spans a one-parameter Abelian group of symmetries 
\cite{myBDlimit}. This property can be used to generate new 
solutions \cite{Dilek}. Here, using the conformal relative of the Roberts 
solution as a seed, we generate the new 
two-parameter family of solutions of vacuum Brans-Dicke gravity
\be
d\hat{s}^2 = \phi^{2\alpha-1} \left[ -dudv +r^2(u,v) d\Omega_{(2)}^2 
\right] \,,
\ee
\be
\hat{\phi} = \phi_0 \left[ \frac{(1-2\sigma)v-u}{(1+2\sigma)v-u 
}\right]^{\pm 2 (1-2\alpha)\sqrt{ \frac{\pi}{|2\omega+3|}} } \,,
\ee
with $r^2$ given by Eq.~(\ref{rsquared}) and the Brans-Dicke coupling by 
Eq.~(\ref{newomega}). These new solutions, labelled by the parameters 
$\sigma$ and $\alpha$ (or, equivalently, $\sigma$ and $\hat{\omega}$) are 
spherically symmetric, time-dependent, and conformal to the Roberts 
spacetime. By repeating the analysis of the previous section, one 
concludes that there is again a central naked singularity where the scalar 
field diverges.

\section{Conclusions}
\setcounter{equation}{0}
\label{sec:5}

The geometry and Brans-Dicke scalar field ~(\ref{new1}), (\ref{rsquared}), 
and~(\ref{Jframescalar} constitute a new solution of vacuum Brans-Dicke 
theory without potential in the Jordan frame. By construction, this 
solution is conformal to the Roberts solution (\ref{Robertsmetric}), 
(\ref{rsquared}), (\ref{Robertsscalar}) of GR with a minimally coupled, 
massless scalar field as the matter source. This new solution has no 
apparent horizons and harbours a central naked singularity, where the 
Ricci curvature ${\cal R}$, the Kretschmann scalar $R_{abcd}R^{abcd}$, and 
the Brans-Dicke scalar $\phi$ diverge. This 
spacetime structure is the same as that of its Einstein frame cousin 
(studied in \cite{Roberts}) used to generate our new solution.

A generalization of the Roberts solution to an Anti-de Sitter 
``background'', conformal to the geometry~(\ref{Robertsmetric}), 
(\ref{rsquared}) has been found by Roberts \cite{RobertsAdS} and is in 
principle useful in studies of the AdS/CFT correspondence. This solution 
is conformal to the $\Lambda=0$ Roberts solution 
\cite{RobertsAdS,Maeda15}.  We do not consider it here because, when 
mapped to the Jordan frame, the negative cosmological constant $\Lambda$ 
of the AdS background gives rise to the negative potential $V(\phi) 
=\Lambda \phi^2/(8\pi) $ \cite{Ali}, as described by 
Eq.~(\ref{ppotential}), and to an imaginary scalar field mass 
($m^2=\frac{\Lambda}{8\pi}$) and is, therefore, of no physical interest.

The fact that the scalar field of the Roberts-AdS spacetime is exactly the 
same as in the Roberts solution without $\Lambda$ has been reported as 
interesting or surprising \cite{RobertsAdS,Maeda15}: in light of the fact 
that a cosmological constant in the Einstein frame corresponds to a mass 
term in the Jordan frame, and that a potential $V(\phi)=m^2\phi^2/2 $ 
disappears completely from the Klein-Gordon equation~(\ref{box}), this 
fact is perhaps not too surprising.  Instead of mapping the 
Roberts-Anti-de Sitter geometry to the Jordan 
frame, we have used a known symmetry of vacuum Brans-Dicke theory to 
generate  a new two-parameter family of solutions. 

Scalar fields in various theories of gravity either collapse to GR black 
holes, as stated by well-known no-hair theorems \cite{nohair}, or they 
seem to generate only naked singularities or wormhole throats.  Outside of 
the context of asymptotically flat and stationary GR or scalar-tensor 
gravity, a general theorem is not available. Future studies will explore 
this direction.\\

\begin{acknowledgments} 
This work is supported, in part, by the Natural 
Sciences \& Engineering Research Council of Canada (Grant no. 2016-03803 
to V.F.) and by Bishop's University. The work of A.G. has been carried out in the framework of 
the activities of the Italian National Group for Mathematical Physics [Gruppo Nazionale per la Fisica
Matematica (GNFM), Istituto Nazionale di Alta Matematica (INdAM)].
\end{acknowledgments}

%\appendix
%\section{ }
%\renewcommand{\theequation}{A.\arabic{equation}}

\end{document}